\documentclass[10pt,english,journal]{IEEEtran}

\usepackage{amsmath,amssymb,amsfonts,mathtools}
\usepackage{amsthm}
\usepackage{bm}
\usepackage{booktabs}
\usepackage{algorithm}
\usepackage[noend]{algpseudocode}
\usepackage{graphicx}
\usepackage{microtype}
\usepackage[hidelinks]{hyperref}
\usepackage[nameinlink]{cleveref}
\crefname{table}{Table}{Tables}
\usepackage{url}
\usepackage{balance}
\usepackage{cite}
\usepackage{enumitem}
\usepackage{siunitx}
\usepackage{verbatim}
\usepackage{caption}
\usepackage{mdframed}
\usepackage[utf8]{inputenc}
\DeclareUnicodeCharacter{2009}{\,}   
\DeclareUnicodeCharacter{2013}{--}   

\theoremstyle{plain}

\theoremstyle{remark}

\usepackage{tabularx}   
\usepackage{array}      
\allowdisplaybreaks 

\newcommand{\SSM}{\mathrm{SSM}}

\usepackage{hyphenat}

\usepackage{bbm}

\usepackage{nicefrac}

\usepackage{float}

\theoremstyle{definition}

\usepackage{adjustbox}
\usepackage{comment}

\usepackage[table]{xcolor}

\makeatletter
\renewcommand{\SSM}{\ifmmode\mathrm{SSM}\else\textsc{SSM}\fi}
\makeatother

\title{Digital Co-Founders: Transforming Imagination into Viable Solo Business via Agentic AI}

\author{
Farhad~Rezazadeh,~\IEEEmembership{Member,~IEEE} and 
Pegah~Bonehgazy

\IEEEcompsocitemizethanks{\IEEEcompsocthanksitem F. Rezazadeh is with the Hostelworld Group, BrainOmega, and the Technical University of Catalonia (UPC) (e-mail: farhad.rezazadeh@upc.edu).}
\IEEEcompsocitemizethanks{\IEEEcompsocthanksitem P. Bonehgazy is with the BrainOmega and the Yazd University.}
\vspace*{-.03cm}
}
\begin{document}
\IEEEaftertitletext{\vspace{-1.5\baselineskip}} 
\maketitle

\begin{abstract}
This paper investigates how individual entrepreneurs can turn creative ideas into successful solo businesses in an era increasingly shaped by Artificial Intelligence (AI) agents. It highlights the key steps that connect personal vision, structured experimentation, and lasting value creation, and shows how AI agents can act as \emph{digital co-founders} throughout this journey. Building on research in entrepreneurship, creativity, and innovation, we present a framework with three key stages: (1) Imagination shaping, where vague goals become clear value propositions, supported by AI agents that help with market scanning, idea refinement, and rapid concept generation; (2) Reality testing, where these ideas are tested through low-cost experiments, structured feedback loops, and efficient execution, with AI agents automating tasks such as prototyping, content creation, customer interaction, and data analysis; and (3) Reality scaling, where successful ideas are transformed into repeatable processes, scalable market strategies, and long-term business models, increasingly operated and optimized by autonomous or semi-autonomous AI workflows. We focus on the specific context of solopreneurship, characterized by limited human resources, complete accountability for decision-making, and a strong association between the founder's identity and the business. The framework clearly identifies key enabling factors such as mental adaptability, effective planning, and successful human-AI collaboration within digital ecosystems. It also thoughtfully addresses ongoing challenges, like uncertainty and cognitive overload, which are heightened by our constant connectivity. 
\end{abstract}


\section{Introduction}

\IEEEPARstart{S}{olo} businesses and one-person ventures are becoming increasingly visible and influential parts of the global economic landscape. Enabled by digital platforms, low-cost tools, and flexible work arrangements, individuals can now conceive, test, and launch products or services without the traditional infrastructure of a larger firm or founding team. At the same time, a new technological layer is emerging. AI agents can autonomously perform tasks, interact with users, and coordinate workflows across tools and platforms. From freelance experts and independent consultants to creators, coaches, micro–Software as a Service (SaaS) founders, and niche e-commerce owners, these \emph{solopreneurs} are not only transforming personal ideas into sustainable economic activity, but increasingly doing so in collaboration with AI systems that function as digital co-workers or even \emph{digital co-founders}. Yet despite the growing relevance of solo, AI-augmented businesses, much of the entrepreneurship literature still centers on startups with human teams, investors, and formal organizational structures. The specific processes by which individuals, supported by AI agents, convert imagination into a viable one-person business remain comparatively underexplored.

At the heart of solo entrepreneurship lies a paradox that AI both complicates and helps resolve. On the one hand, solo founders enjoy a high degree of freedom. They can define their own vision, choose their own pace, and pivot without negotiation. On the other hand, this same freedom is constrained by limited human resources, cognitive bandwidth, and emotional resilience. A single individual must simultaneously serve as strategist, creator, marketer, operator, and financial manager. AI agents can partially absorb some of these roles—handling routine tasks, running experiments, or interacting with customers—but they also introduce new demands: prompt design, system oversight, ethical judgment, and strategic orchestration of human–AI collaboration. 

Traditional models of entrepreneurship and innovation—often developed in the context of larger firms or human founding teams—only partially capture this new configuration. There is a need for a more precise understanding of how imagination is structured, tested, and scaled under the specific conditions of solo business creation in an AI-agentic \cite{rezazadeh2025agentic} environment.
Imagination occupies a central place in this process. It is not merely the source of creative ideas, but the cognitive space in which individuals simulate possible futures, envision different identities, and explore alternative value propositions. For solo founders, imagination is deeply intertwined with self-concept; business ideas are frequently extensions of personal interests, values, or life experiences. AI agents expand this imaginative space by rapidly generating alternatives, scenarios, and narratives, surfacing patterns in data, and exposing founders to a broader range of possibilities than they could easily produce on their own. However, imagination—human or AI-augmented—remains insufficient on its own. The challenge is to systematically transform a fluid, often ambiguous inner world, enriched but also potentially overwhelmed by AI-generated options, into concrete offers that solve specific problems for clearly defined audiences. This transformation—\emph{turning imagination into reality}—requires a blend of creativity, disciplined experimentation, operational structure, and effective human–AI collaboration.

For a solo founder, a pivot is not just a strategic move; it can feel like a shift in personal identity and in the perceived role of AI in their work.
This paper addresses this gap by focusing specifically on the mechanisms through which solo entrepreneurs—supported by AI agents—convert imaginative ideas into viable businesses. We propose a conceptual framework structured around three core stages: i) imagination shaping, ii) reality testing, and iii) reality scaling, each involving an evolving configuration of human and AI agency.

Focusing on AI-augmented solo businesses reveals several distinctive challenges and enablers within this three-stage process. Limited financial and human resources still require solo founders to maximize learning per unit of time, energy, and money, but AI agents can dramatically increase their effective capacity. Decision-making speed is high, yet the constant availability of AI-generated options and recommendations may amplify decision fatigue. Emotional factors—such as fear of visibility, impostor syndrome, and risk aversion—interact with new concerns about dependence on AI, loss of authenticity, or uncertainty about the reliability and transparency of agentic systems. At the same time, solopreneurs can leverage infrastructures that were not widely available in earlier eras; digital platforms for distribution, no-code and low-code tools for building products, AI agents that orchestrate content and workflows across tools, and online communities for feedback and collaboration. These elements effectively extend the capabilities of a one-person business, making the imagination-to-reality pipeline more accessible, more automated, and potentially more scalable.

\subsection{Related Work}
\label{sec:relatedwork}

Research on solo business operates at the intersection of solo self-employment, entrepreneurial imagination and creativity, process-oriented approaches to opportunity development, and, increasingly, the role of AI in entrepreneurship. In this section, we synthesize four main streams: (1) the socio-economic and psychological characteristics of solo self-employment, (2) imagination and creativity in entrepreneurship, (3) process models such as effectuation, lean startup, and design thinking that structure the transition from idea to implementation, and (4) emerging work on AI-enabled and agentic forms of individual entrepreneurship.

\subsubsection{Solo self-employment and One-person Enterprises}

A first stream examines solo self-employment and one-person enterprises as distinct, increasingly prevalent forms of entrepreneurship. Belt's work on one-person enterprises documents how ``solopreneurs'' constitute a nontrivial share of the workforce and explores their relationship to business growth and personal well-being \cite{Belt2015}. Cie\'slik and Van Stel conceptualize solo self-employment as a ``third segment'' in the labour market (alongside employees and employers), highlighting key policy challenges around decent work, social protection, activation of marginalized groups, and well-being \cite{CieslikVanStel2023}. Building on this foundation, Cie\'slik \cite{Cieslik2025Expansion} segments solo self-employed workers by job characteristics and expansion strategies, emphasizing heterogeneity between necessity-based solos and highly skilled professionals who view solo status as a stepping stone toward employer entrepreneurship. Related labour-market analyses from the International Labour Organization (ILO) and European agencies show that solo self-employment is both numerically dominant within self-employment and often linked to constrained wage-employment alternatives in specific occupations and countries \cite{Mitra2022}. Broader studies connect solo self-employment to the rise of hybrid and part-time entrepreneurship, highlighting the interaction between solo work, upskilling, and changing job compositions across Europe and beyond \cite{Bogenhold2017, Kana2022}.

A second subset of this literature looks at the quality of work, identity, and resilience among solo entrepreneurs. Holloway analyzes solo self-employment as an economically marginal yet numerically dominant form of work, connecting it to broader questions of entrepreneurial subjectivity and spatial precarity \cite{Holloway2021}. Van den Groenendaal and colleagues qualitatively examine career self-management strategies and career sustainability among solo self-employed workers, identifying patterns of enablers and barriers over time \cite{vandenGroenendaal2022}. Schummer \cite{Schummer2019} studies need satisfaction (autonomy, competence, relatedness) and affective commitment, showing that psychological needs explain commitment differently for solo self-employed versus employer entrepreneurs. In parallel, work on solopreneur resilience during crises, such as Ebrahimi's Crisis-Responsive Solopreneur Resilience Model (CRSRM), highlights the role of intrinsic motivation, continuous learning, professional networks, and social support in maintaining adaptive capacity under uncertainty \cite{Ebrahimi2025}. Other contributions emphasize the downside risks of solopreneurship, including low survival rates of small and Medium-sized Enterprises (SMEs) in specific regions \cite{Kana2022} and ambiguous boundaries between independent and dependent work in the gig economy \cite{Bogenhold2017}.
\subsubsection{Entrepreneurial imagination and creativity}

A growing body of work addresses entrepreneurial imagination and creativity, providing a foundation for understanding how individuals turn \emph{what could be} into \emph{what is}. Lecuna explicitly examines imagination as a complementary condition to discovery and creation theories in entrepreneurship, arguing that entrepreneurial imagination has been underdefined and undermeasured in prior research \cite{Lecuna2024}. Kier's work on entrepreneurial imaginativeness in new ventures conceptualizes imagination as a forward-looking mechanism that enables entrepreneurs to construct novel combinations of resources and futures \cite{Kier2018}. Thompson's critique and renewal of the entrepreneurial imagination further positions imagination as central to how entrepreneurs reconfigure knowledge and envision new possibilities \cite{Thompson2018}. Jones \cite{Jones2015} explores how appropriability concerns and historical legacies shape entrepreneurial imagination in cross-border contexts. 

Parallel research on creativity in entrepreneurship reinforces this view. Fillis and Rentschler review the role of creativity in entrepreneurship, framing it as a holistic and transdisciplinary construct that permeates opportunity recognition, innovation, and venture growth \cite{FillisRentschler2010}. Santos et al.\ \cite{Santos2019} provide an overview of creativity in the context of creative industries, distinguishing creative entrepreneurs from more conventional business founders. More recent quantitative work shows how creativity and entrepreneurial alertness interact to channel novel ideas toward more promising opportunities and firm performance, underscoring that creativity must be coupled with opportunity filters to yield economic value \cite{Karami2024}. These contributions suggest that imagination and creativity are not merely initial sparks but ongoing capabilities that shape how solo founders continuously reimagine their businesses and identities.

\subsubsection{Process-based Theories of Opportunity Development}

A third stream consists of process-oriented theories that describe how entrepreneurs move from ideas to implemented ventures. Sarasvathy's theory of effectuation distinguishes between causation (goal-driven, predictive planning) and effectuation (means-driven, adaptive action) as alternative logics in entrepreneurial decision-making under uncertainty \cite{Sarasvathy2001}. Effectuation emphasizes starting from available means, building partnerships, and co-creating opportunities, which is particularly relevant for solo entrepreneurs operating with limited resources and high autonomy. The lean startup methodology popularized by Ries operationalizes iterative ``build--measure--learn'' cycles centered on Minimum Viable Products (MVPs), validated learning, and continuous innovation \cite{Ries2011}. Brown's work on design thinking in innovation and strategy adds a human-centered, iterative approach that moves from inspiration through ideation to implementation, using prototyping and user feedback as core mechanisms for reducing uncertainty \cite{Brown2008}. Taken together, these frameworks provide structured approaches for reality testing and reality scaling, even though they were not originally developed with one-person ventures as the primary context.

\subsubsection{AI agents and Entrepreneurship}

A fourth, rapidly expanding stream addresses AI and entrepreneurship, including the role of AI agents and AI-based technologies in reshaping entrepreneurial processes. Early conceptual work framed AI as an external enabler that reshapes uncertainty, opportunity structures, and venture creation in the ``Fourth Industrial Revolution'' \cite{Chalmers2021}. Subsequent systematic reviews and surveys synthesize this emerging field: Giuggioli and Pellegrini conceptualize AI as an enabler for entrepreneurs, emphasizing enhanced decision-making, new opportunity spaces, and AI-supported entrepreneurial education \cite{GiuggioliPellegrini2023}; Fossen et al.\ provide an extensive survey of how AI affects opportunity creation, decision-making, entrepreneurial entry, and ecosystems, including evidence that entrepreneurs---and particularly the self-employed---are among the most intensive users of AI tools \cite{Fossen2024}; Uriarte \cite{Uriarte2025} offers a hybrid systematic review of AI-based technologies in entrepreneurship, mapping research clusters around AI-enabled business model innovation and contextual factors. Al-Mamary \cite{AlMamary2025} develops a conceptual model of AI capabilities (decision-making, automation, customer experience, innovation, and risk mitigation) for the success of entrepreneurial ventures. Most pertinent to the present paper, Ganuthula's AI-enabled Individual Entrepreneurship Theory (AIET) argues that AI fundamentally transforms individual entrepreneurial capacity by augmenting skills, restructuring capital access, and reshaping risk profiles, thereby challenging traditional assumptions about scale and resource dependence \cite{Ganuthula2025}. 

While this work primarily treats AI at the level of ``individual entrepreneurs'' in general, it offers firm theoretical grounding for understanding how AI tools---including increasingly agentic systems---can expand what one-person ventures can do without building traditional organizations. At the same time, the literature has only begun to explore how AI tools, and especially semi-autonomous AI agents, reconfigure the micro-processes of solo venture creation. Most existing contributions examine AI adoption at the firm, ecosystem, or undifferentiated ``entrepreneurs,'' level, rather than the specific challenges and identity dynamics of solo founders leveraging AI agents as enduring collaborators. This gap motivates the framework developed in the present paper, which focuses on how AI-augmented solopreneurs turn imagination into reality through staged processes of imagination shaping, reality testing, and reality scaling.

\subsection{Contributions}

Building on the four streams of literature reviewed in Section~\ref{sec:relatedwork}---solo self-employment and one-person enterprises, entrepreneurial imagination and creativity, process-based theories of opportunity development, and AI agents in entrepreneurship---this paper makes four interrelated contributions. Together, they address four main gaps that emerge from the related work: (Gap~1) the dominance of structural and descriptive accounts of solo self-employment over micro-process views, (Gap~2) the lack of imagination- and creativity-focused work that is specific to solopreneurs, (Gap~3) the limited adaptation of process frameworks (effectuation, lean startup, design thinking) to one-person ventures, and (Gap~4) the nascent state of AI-and-entrepreneurship research at the level of solo, AI-augmented micro-processes.

\subsubsection{Conceptualizing AI-augmented Solo business as a Distinct Imagination-to-Reality Context}

First, we reframe solo business---including AI-augmented solo business---as a distinct entrepreneurial context in which specific structural, psychological, and technological conditions shape the path from imagination to implementation. In AI-augmented solo ventures, founders face:

\begin{itemize}
    \item extremely limited human and financial resources but expanded ``computational'' capacity through AI tools and agents,
    \item heightened role accumulation (one person as strategist, creator, operator, brand, and orchestrator of AI agents), and
    \item a strong coupling between founder identity, lifestyle choices, business model, and the personalized AI ecosystem that supports their work.
\end{itemize}

Where much of the existing solo self-employment literature remains descriptive and structural (segmenting types of solo workers, documenting labour-market outcomes, or discussing policy challenges), our work shifts the focus to micro-level processes; how individual solopreneurs actually work with their ideas over time and how AI agents participate in those processes. By placing imagination at the center of this AI-augmented solo context, we show that solo entrepreneurship is not merely ``a smaller startup'' or ``employment of last resort'' but a distinct way of turning personal visions into economic activity with the assistance of semi-autonomous digital collaborators.

\begin{mdframed}[
  backgroundcolor=gray!5,
  linecolor=gray!50,
  linewidth=0.5pt,
  roundcorner=2pt,
  innerleftmargin=8pt,
  innerrightmargin=8pt,
  innertopmargin=6pt,
  innerbottommargin=6pt,
  skipabove=6pt,
  skipbelow=6pt,
  userdefinedwidth=\columnwidth,
  align=center
]
\centering
\footnotesize
\textbf{Scope of This Contribution}\\[2pt]
This contribution directly addresses Gap~1 by providing a process-oriented,
psychologically grounded, and technologically informed view that complements
existing macro- and policy-oriented analyses of solo self-employment, and it
also begins to respond to Gap~4 by explicitly situating AI within the solo
context.
\end{mdframed}


\begin{figure*}[t]
\centering
\includegraphics[width=\linewidth]{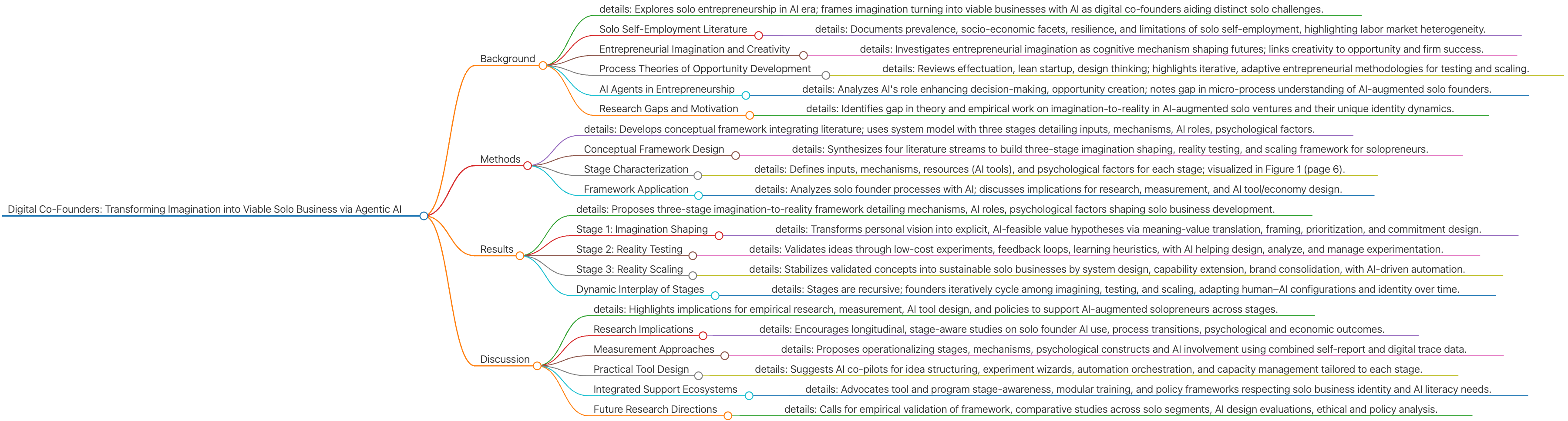}
\caption{The figure summarizes the background literatures, the three-stage imagination-to-reality framework (imagination shaping, reality testing, reality scaling), and the resulting research and practical implications for AI-augmented solo entrepreneurship.}

\label{fig:mindmap-paper}
\end{figure*}

\subsubsection{Integrating Imagination, Creativity, and Human--AI Collaboration in the Lived Realities of Solopreneurs}

Second, we contribute to the literature on entrepreneurial imagination and creativity by embedding these constructs within the concrete realities of solo work under pervasive AI assistance. Prior research conceptualizes imagination as a forward-looking capability that allows entrepreneurs to envision possibilities, recombine resources, and construct new futures, but typically treats ``the entrepreneur'' in a generic way and assumes predominantly human-only cognition. Our framework shows that in AI-augmented solo businesses:

\begin{itemize}
    \item imagination is tightly linked to identity (the business is often an extension of the person’s skills, values, life story, and preferred forms of human--AI collaboration);
    \item imaginative processes are constrained and shaped by cognitive bandwidth, emotional regulation, and the absence of a founding team to provide complementary mental models---with AI agents partially compensating but also introducing new demands of oversight, judgment, and alignment; and
    \item creativity must be continuously balanced with economic necessity, self-protection, and well-being (e.g., avoiding burnout, managing risk exposure, and sustaining motivation when working alone but surrounded by always-available AI tools).
\end{itemize}

We introduce the notion of \emph{imagination shaping} as a distinct stage; the disciplined transformation of diffuse personal visions into explicit value hypotheses, under solo-specific and AI-specific constraints. This stage highlights mechanisms that imagination research has largely treated implicitly, such as:

\begin{itemize}
    \item translating personal meaning into market-relevant value propositions with the support of generative and analytical AI,
    \item managing inner multiplicity (many possible futures) when there is no human team to negotiate trade-offs but a potentially overwhelming set of AI-generated options, and
    \item using small, ``identity-safe'' and low-cost experiments---often designed, executed, or monitored by AI agents---to test ideas without threatening self-concept.
\end{itemize}

In doing so, we bridge the gap between abstract theories of entrepreneurial imagination/creativity and the lived experience of solopreneurs, spelling out how imagination functions when one person carries both the idea and its full execution, while coordinating an evolving ensemble of AI agents.
\begin{mdframed}[
  backgroundcolor=gray!5,
  linecolor=gray!50,
  linewidth=0.5pt,
  roundcorner=2pt,
  innerleftmargin=8pt,
  innerrightmargin=8pt,
  innertopmargin=6pt,
  innerbottommargin=6pt,
  skipabove=6pt,
  skipbelow=6pt,
  userdefinedwidth=\columnwidth,
  align=center
]
\centering
\footnotesize
\textbf{Scope of This Contribution}\\[2pt]
This contribution addresses Gap~2 by providing a solo-specific, psychologically
grounded account of entrepreneurial imagination and creativity, and also
contributes to Gap~4 by explicitly incorporating human--AI collaboration into
imaginative processes.
\end{mdframed}


\subsubsection{Adapting Process Frameworks to AI-agentic Solo Business Workflows}

Third, we extend and adapt existing process frameworks---effectuation, lean startup, and design thinking---to the particular conditions of AI-enabled one-person ventures. These approaches already provide powerful logics for dealing with uncertainty. Still, they were developed with human teams, organizational routines, or external investors implicitly in mind and generally without explicit consideration of AI agents as ongoing actors in the process.

Our framework reorganizes and updates them into three solo-focused, AI-aware stages that align with the opportunity development literature:

\paragraph{Imagination Shaping} 
Effectual logic is reinterpreted as starting from one’s means (skills, identity, network, and accessible AI tools/agents) while explicitly treating inner vision as a resource to be structured. Design thinking informs the reframing of personal interests into problem statements grounded in real users, supported by AI for market scanning and sensemaking. A lean mindset guides the articulation of initial value hypotheses that can be tested with minimal human effort, often through AI-assisted ideation and rapid content or prototype generation.

\paragraph{Reality Testing}
Lean startup principles are adapted to translate value hypotheses into micro--MVPs and ``nano experiments'' that fit a solo founder’s time and resource constraints (e.g., pre-selling, simple landing pages, automated email sequences, or AI-assisted A/B tests). Design thinking emphasizes rapid, low-fidelity prototyping and customer conversations, recognizing that the same individual---supported by AI agents for drafting, scheduling, and analysis---designs, implements, and interprets feedback. Effectuation’s notions of affordable loss and partnership-building are revisited in a context where available partners include not only clients, collaborators, and platforms, but also off-the-shelf or custom AI agents integrated into workflows.

\paragraph{Reality Scaling}
In the scaling stage, we adapt lean and effectual principles to stabilize routines (e.g., simple operating systems, habit loops, and codified processes) and then partially automate them via AI agents rather than building formal organizations. Here, scale is pursued through capability extension via tools (automation, multi-agent orchestration, no-code platforms, and digital infrastructure) rather than hiring employees, offering a structurally different path to growth for solo businesses. Our framework clarifies which parts of effectuation, lean startup, and design thinking must be reinterpreted when there is no human team or investor pressure. Still, there is a growing ``team'' of AI agents whose behavior must be designed, monitored, and aligned.
\begin{mdframed}[
  backgroundcolor=gray!5,
  linecolor=gray!50,
  linewidth=0.5pt,
  roundcorner=2pt,
  innerleftmargin=8pt,
  innerrightmargin=8pt,
  innertopmargin=6pt,
  innerbottommargin=6pt,
  skipabove=6pt,
  skipbelow=6pt,
  userdefinedwidth=\columnwidth,
  align=center
]
\centering
\footnotesize
\textbf{Scope of This Contribution}\\[2pt]
This contribution directly addresses Gap~3 by tailoring the major process
frameworks to the solo context, and further contributes to Gap~4 by specifying
how AI agents are embedded within these adapted stages of opportunity
development.
\end{mdframed}


\subsubsection{Bridging Macro Descriptions, Micro Mechanisms, and AI Design Implications}

Finally, our framework connects macro-level and micro-level perspectives and adds explicit design implications for AI and tool ecosystems. On the one hand, it is grounded in macro-level observations on the prevalence, heterogeneity, and policy significance of solo self-employment, as well as the growing diffusion of AI tools among self-employed workers. On the other hand, it specifies micro mechanisms---cognitive, emotional, behavioral, and technological---that explain how AI-augmented solopreneurs actually turn imagination into reality. This bridging contribution enables:

\begin{itemize}
    \item empirical work that can operationalize our three stages (imagination shaping, reality testing, and reality scaling) as measurable constructs, link them to performance and well-being outcomes, and examine their presence across different segments of solo workers (e.g., necessity vs.\ opportunity solopreneurs, creative professionals vs.\ technical micro--SaaS founders, low vs.\ high AI adoption);
    \item studies that explicitly model the configuration and role of AI agents at each stage (e.g., which tasks are delegated, how agency is shared, and how human oversight is exercised); and
    \item practical interventions (training, coaching, AI agents, and support programs) that are tailored to the specific frictions in each stage---for example, tools that help structure and prioritize ideas for solo founders overwhelmed by possibilities, reduce the perceived emotional risk of early experiments through low-exposure testing environments, or automate routine tasks via agents to free cognitive resources for continued imaginative and strategic work.
\end{itemize}
\begin{mdframed}[
  backgroundcolor=gray!5,
  linecolor=gray!50,
  linewidth=0.5pt,
  roundcorner=2pt,
  innerleftmargin=8pt,
  innerrightmargin=8pt,
  innertopmargin=6pt,
  innerbottommargin=6pt,
  skipabove=6pt,
  skipbelow=6pt,
  userdefinedwidth=\columnwidth,
  align=center
]
\centering
\footnotesize
\textbf{Scope of This Contribution}\\[2pt]
This contribution consolidates the response to Gap~1 (by linking macro
descriptions of the solo economy to micro mechanisms) and Gap~4 (by turning
high-level claims about AI in entrepreneurship into concrete design
implications for AI agents and ecosystems in solo ventures).
\end{mdframed}



The remainder of the paper is structured as follows. \Cref{sec:framework} presents our conceptual framework, detailing the three
stages of imagination shaping, reality testing, and reality scaling, and
specifying the mechanisms, human–AI configurations, resources, and
psychological factors involved in each.
\Cref{sec:implications} discusses implications for research and practice,
including potential measurement approaches and design principles for AI agents
and tool ecosystems that support solopreneurs.
\Cref{sec:conclusion} concludes by summarizing the key insights and suggesting
avenues for future work on turning imagination into reality in AI-enabled solo
business contexts.
Figure \ref{fig:mindmap-paper} shows a mindmap overview of the paper.

\begin{figure*}[t]
\centering
\includegraphics[width=\linewidth]{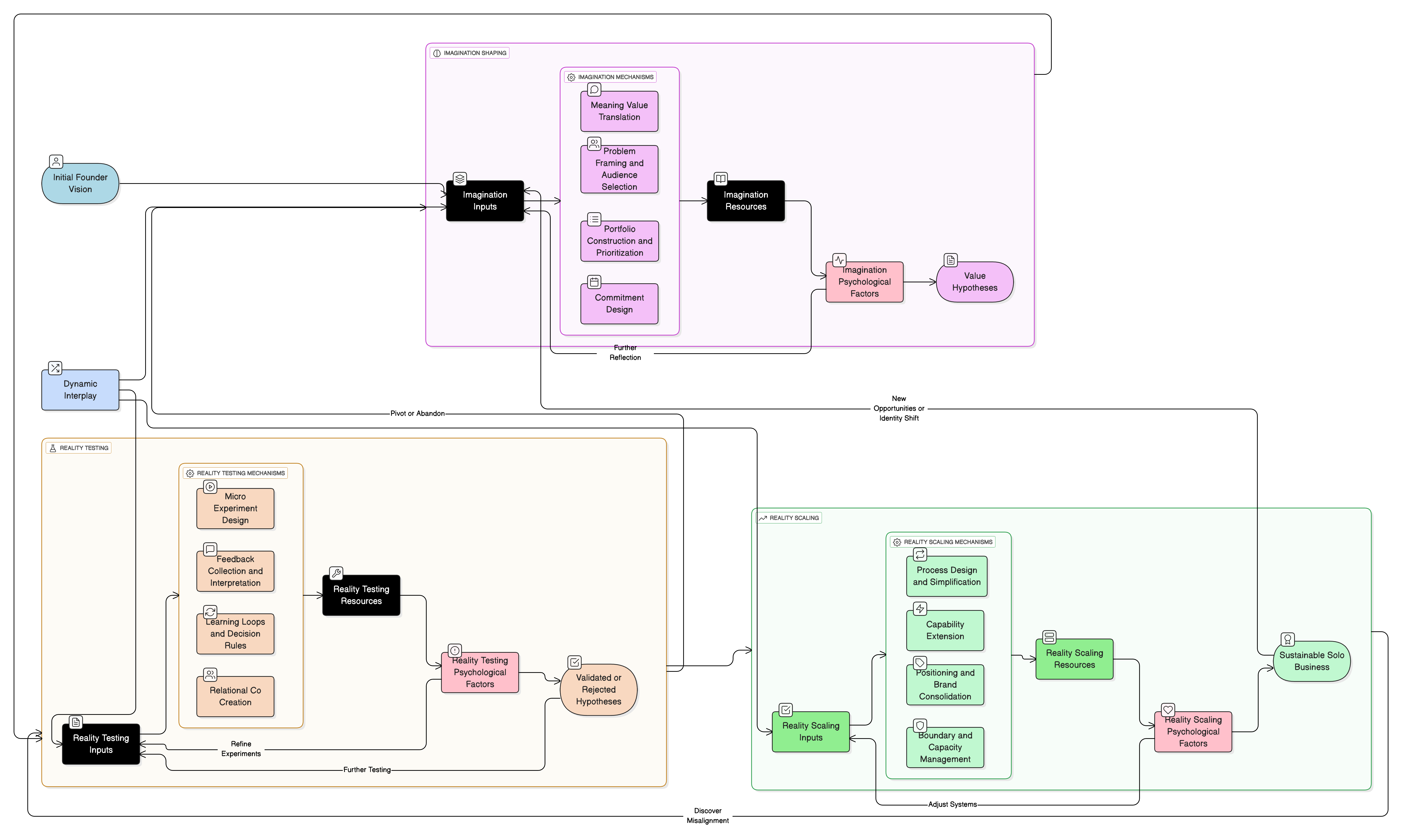}
\caption{Conceptual three-stage framework of how solo founders move from an initial founder vision to a sustainable solo business. The model distinguishes imagination shaping, reality testing, and reality scaling, and, within each stage, highlights the role of inputs, mechanisms, resources (including digital and AI-based tools), and psychological factors. Arrows and feedback loops capture the dynamic interplay over time, including further reflection, refined experiments, system adjustments, pivots, and the emergence of new opportunities or identity shifts.}

\label{fig:system-modeling}
\end{figure*}

\section{Conceptual Framework: From Imagination to AI-Augmented Solo Business Reality}
\label{sec:framework}

This section presents our conceptual framework for understanding how solo founders, supported by AI tools and agents, transform inner visions into viable one-person businesses. As shown in Figure \ref{fig:system-modeling}, the framework is organized into three interrelated stages: \emph{imagination shaping}, \emph{reality testing}, and \emph{reality scaling}. Rather than a linear pipeline, these stages form a recursive system: AI-augmented solopreneurs often move back and forth between them as they learn, pivot, reconfigure their human--AI setup, and renegotiate their identity and goals. Each stage is characterized by specific inputs, mechanisms, resources (including AI agents), and psychological factors, which we outline below.

\subsection{Overview of the Three Stages}

\emph{Imagination shaping} focuses on transforming diffuse, personally meaningful ideas into explicit, testable value hypotheses. Here, the solopreneur works primarily \emph{in the mind}, but in a disciplined way; clarifying who the business is for, what problem it might solve, how it relates to the founder’s own skills, values, and aspirations, and how AI tools and agents might expand what is feasible for one person.

\emph{Reality testing} centers on exposing these value hypotheses to the outside world through low-cost experiments. The emphasis is on learning under uncertainty: gathering feedback from potential users, refining offers, and making decisions about which directions to pursue, pivot, or abandon. AI agents support this stage by accelerating experiment design, content creation, data collection, and analysis.

\emph{Reality scaling} addresses how validated concepts are stabilized and expanded into a sustainable solo business. This stage involves designing simple systems, routines, and supporting infrastructures that allow the founder to deliver value repeatedly without exceeding personal and resource constraints. AI agents increasingly act as persistent, semi-autonomous workflows that help operate and optimize the business.

Across all three stages, the founder’s identity, emotional state, and cognitive bandwidth interact with available tools (including AI), networks, and market conditions, shaping what is imagined, what is tried, and what ultimately becomes real.

\subsection{Stage 1: Imagination Shaping}
Imagination shaping is the process by which AI-augmented solopreneurs convert inner visions into structured value hypotheses. It sits at the intersection of personal meaning, market relevance, and perceived human--AI capabilities.

\subsubsection{Inputs}
Typical inputs to this stage include the founder’s prior experiences, skills, and interests, which filter what feels meaningful and credible for them to pursue. Imagination shaping is also influenced by observed frustrations or unmet needs in their environment that suggest where new value might be created, as well as by exposure to role models and narratives about solo and AI-enabled businesses that expand (or narrow) their sense of what kinds of ventures are possible. At the same time, practical constraints such as time, money, location, or family responsibilities delineate the feasible space of action, while the set of easily accessible digital tools and AI systems (e.g., chatbots, specialized agents, and no-code platforms) shapes what seems ``doable'' for one person. Together, these elements form a raw ``imaginative field’’ in which many possible futures coexist, some of which are newly enabled by AI.

\subsubsection{Core Mechanisms}
We highlight four mechanisms at work in imagination shaping, each of which can be amplified or structured by AI agents:

\textbf{\emph{Meaning--value translation.}}
The founder asks: ``What matters to me, and how might that matter to someone else?'' Personally meaningful themes (e.g., creativity, freedom, social impact, and technical mastery) are translated into candidate value propositions (e.g., a design studio, a coaching practice, and 
a niche SaaS tool). Generative AI can assist by suggesting alternative framings, audiences, and narratives, effectively expanding the space of imagined offers while the human founder judges coherence and fit.

\textbf{\emph{Problem framing and audience selection.}}
Diffuse ideas are reframed as problem--solution hypotheses:
\begin{itemize}
    \item Who might benefit?
    \item What pain, aspiration, or job-to-be-done is addressed?
    \item Why would this be preferable to alternatives?
\end{itemize}
AI tools can support this mechanism through lightweight market scans (e.g., summarizing competitor offerings and surfacing common pain points in niche communities) and by helping articulate positioning statements. At this point, the goal is not precision but coherent direction.

\textbf{\emph{Portfolio construction and prioritization.}}
Many solopreneurs initially generate a portfolio of imagined businesses. Imagination shaping involves ranking these options using simple criteria; perceived fit with identity, feasible starting steps, required resources (including AI capabilities), and risk exposure. Here, AI agents can help structure comparison tables, simulate basic scenarios, or highlight which options align best with the founder’s stated values and constraints. The result is a shortlist of experiments, rather than a single fixed plan.

\textbf{\emph{Commitment design.}}
Finally, the founder defines small, time-bound commitments (e.g., ``talk to five potential clients'', ``launch a landing page'', and ``deploy a basic AI assistant for one workflow'') that move the idea from thought to action. These commitments are intentionally limited to preserve emotional safety and flexibility. AI can further reduce perceived risk by lowering the cost of producing early artefacts (draft copy, mockups, and simple automations).

\subsubsection{Resources}
Key resources at this stage include reflective practices (journaling, mind-mapping), structured prompts (business model canvases, value proposition canvases), exposure to communities and mentors, and digital tools that help articulate and refine ideas. General-purpose and domain-specific AI systems act as cognitive amplifiers. They generate examples, rephrase ideas for different audiences, and perform rapid, approximate desk research that supports early framing.

\subsubsection{Psychological Factors} Imagination shaping is heavily influenced by:
\begin{itemize}
    \item \textbf{Identity clarity:} how the founder sees themselves and their desired future self, including their comfort with integrating AI into their work and self-story;
    \item \textbf{Tolerance for ambiguity:} willingness to hold multiple possible futures---both human- and AI-intensive---without premature closure;
    \item \textbf{Self-efficacy and perceived legitimacy:} beliefs about ``being the kind of person'' who can run such a business and effectively direct AI agents;
    \item \textbf{Emotional safety:} fear of visibility or judgment (of both human work and AI-assisted work) can lead to keeping ideas purely private, preventing transition to reality testing.
\end{itemize}

The output of this stage is a set of explicit, personally meaningful, AI-feasible, and plausibly actionable value hypotheses that can be brought into contact with the real world.

\subsection{Stage 2: Reality testing}

Reality testing is where imagined futures collide with actual users, markets, and constraints. The emphasis shifts from internal coherence to external validity and learning, with AI agents increasingly used to lower the cost and friction of experimentation.

\subsubsection{Inputs}
This stage takes as input the value hypotheses and small commitments generated in imagination shaping; tentative definitions of target audiences, problems, possible offers, and initial ideas about how AI might support delivery or operations.

\subsubsection{Core Mechanisms} This stage consists of four core mechanisms that structure how solopreneurs expose their ideas to the real world and learn from the outcomes:

\textbf{\emph{Micro-experiment design.}} Solopreneurs convert hypotheses into minimal experiments that are affordable in terms of time, money, and emotional exposure, such as:
validation activities at this stage can include informal interviews or problem-discovery calls to better understand prospective users’ needs, low-fidelity prototypes or mockups that make the concept more tangible, and simple landing pages or social media posts that gauge initial interest. They can also involve pre-selling or running pilot programs with a handful of early adopters to test willingness to pay and real-world usage, as well as small-scale deployments of AI agents (e.g., a support chatbot or a content-generation assistant) to evaluate specific aspects of the envisioned service.

These ``nano MVPs'' are smaller and more personal than the classic startup MVP, reflecting the limited resources and close identity involvement of solo founders. AI tools and agents assist by generating copy, designing visuals, building basic automations, and instrumenting experiments with simple tracking.

\textbf{\emph{Feedback collection and interpretation.}}
The founder gathers data (qualitative and/or quantitative) on user reactions, willingness to pay, perceived value, and fit with their own preferences. AI can help transcribe, summarize, and cluster feedback, or run basic analyses on click-through or conversion data. Interpretation, however, remains deeply human and is shaped by:
\begin{itemize}
    \item confirmation bias (hearing what confirms the original vision);
    \item emotional defensiveness (taking criticism as personal, especially when the offer is closely tied to identity);
    \item prior identity and AI commitments (reluctance to pivot away from self-defining ideas or to abandon a chosen AI-centric approach).
\end{itemize}

\textbf{\emph{Learning loops and decision rules.}}
Reality testing is organized around repeated build--measure--learn or try--sense--adapt loops. The founder updates their mental model and decides whether to:
\begin{itemize}
    \item persevere (deepen the current direction),
    \item pivot (adjust audience, positioning, channel, offer, or the role of AI), or
    \item park/abandon (suspend an idea and return to imagination shaping).
\end{itemize}
These decisions are often governed by simple heuristics such as ``affordable loss'', ``energy gain vs.\ energy drain'', or ``signal of demand vs.\ silence''. AI agents can support these loops by maintaining experiment logs, surfacing patterns over time, and prompting the founder to revisit or reframe hypotheses.

\textbf{\emph{Relational co-creation.}}
Particularly for solo businesses, early clients, peers, and collaborators often function as co-designers. Offers are refined in ongoing relationships, blurring the line between market research and service delivery. AI can mediate aspects of these relationships (e.g., automated follow-ups, collaborative whiteboards, and co-created content), but also raises new questions about authenticity and transparency that the founder must navigate.

\subsubsection{Resources}
Important resources include access to potential customers (offline or online communities, platforms), low-cost digital tools for prototyping and distribution, and peer or mentor feedback that helps interpret noisy signals. AI-based tools---for survey design, analytics, content generation, and workflow automation---extend what a single person can experiment with in a limited time.

\subsubsection{Psychological Factors}
Reality testing involves exposure and risk, and is strongly affected by:
\begin{itemize}
    \item \textbf{Fear of rejection and visibility:} reluctance to ``go public'' can delay or dilute experiments; the ease of generating polished AI content may both lower and raise this fear;
    \item \textbf{Resilience:} capacity to absorb negative feedback or failed experiments without abandoning the broader vision;
    \item \textbf{Learning orientation:} framing outcomes as information rather than verdicts on self-worth or on the legitimacy of AI support;
    \item \textbf{Decision fatigue:} repeated small choices made alone (including which AI tools to use and how to configure them) can deplete cognitive and emotional resources.
\end{itemize}

The output of this stage is a set of validated, refined, or rejected hypotheses, along with experiential learning about both the market and the founder’s own preferences, capabilities, and effective human--AI arrangements.

\subsection{Stage 3: Reality scaling}

Reality scaling concerns the transition from sporadic, experiment-driven activity to a more stable and sustainable solo business, operated increasingly through systems and AI-supported workflows. The goal is to create sufficient structure to support growth and reliability, without imposing bureaucratic overhead that exceeds one person’s capacity.

\subsubsection{Inputs}
This stage starts from what has proven viable during reality testing; validated value propositions, early paying clients, repeatable activities that generate results, functioning AI-assisted workflows, and a clearer sense of the founder’s preferred working style and tolerance for automation.

\subsubsection{Core Mechanisms}

\emph{Process design and simplification.}
The founder identifies recurring tasks (e.g., client acquisition, onboarding, delivery, billing, and content production) and turns them into simple, documented routines: checklists, templates, scripts, and calendar blocks. The emphasis is on minimum viable systems that reduce cognitive load and variability. AI agents can be embedded into these routines as assistants (e.g., drafting emails), monitors (e.g., flagging anomalies), or semi-autonomous actors (e.g., handling standard support queries).

\textbf{\emph{Capability extension through tools, agents, and partnerships.}}
Instead of building a team, many solopreneurs scale by extending their capabilities through:
\begin{itemize}
    \item automation, no-code tools, and multi-agent AI orchestrations;
    \item AI systems for content, analytics, product generation, or operations;
    \item freelancers, platforms, and strategic partnerships that provide narrow, on-demand expertise.
\end{itemize}
This produces a networked and agentic form of scaling, where the founder remains the central node, coordinating a constellation of tools, AI agents, and human collaborators rather than managing employees.

\textbf{\emph{Positioning and brand consolidation.}}
Reality scaling also involves stabilizing how the business is perceived; sharpening niche definition, refining messaging, and building a recognizable personal or product brand. AI tools can support consistent content and presence across channels, while the founder curates tone and authenticity. Clear positioning supports trust, pricing power, and more predictable demand.

\textbf{\emph{Boundary and capacity management.}}
Because the business and the person are tightly coupled, scaling requires deliberate boundary-setting; choosing what not to do, setting working hours, defining client fit criteria, and protecting time for continued imaginative work. AI can help enforce boundaries (e.g., scheduling rules and inbox triage) but may also tempt over-extension by making ``just one more project'' technically feasible. Managing this tension is central to sustainable solo scaling.

\subsubsection{Resources}
Key resources include digital infrastructure (payment systems, Customer Relationship Management (CRM), scheduling, and learning platforms), standard operating procedures, AI and automation stacks, and a supportive ecosystem (communities, mentors, professional networks) that facilitates referrals, knowledge sharing, and collaboration.

\subsubsection{Psychological Factors}
Reality scaling involves a shift from exploratory experimentation to ongoing stewardship of a living system. It is shaped by:
\begin{itemize}
    \item \textbf{Identity consolidation:} ``This is who I am, what I’m known for, and how AI fits into my practice'';
    \item \textbf{Motivational balance} between novelty-seeking and tolerance for routine and systematization;
    \item \textbf{Long-term resilience}, including habits around rest, learning, and financial management in a partly automated environment;
    \item \textbf{Willingness to let go} of some imaginative possibilities and some manual control in order to focus on what works and delegate to tools and agents where appropriate.
\end{itemize}

The output of this stage is a sustainable one-person business. A configuration of routines, relationships, tools, and AI agents that reliably converts the founder’s capabilities into value for others, while maintaining alignment with their desired identity and life design.

\subsection{Dynamic interplay between stages}

Although analytically distinct, the three stages are interdependent and recursive. Insights from reality testing may trigger new cycles of imagination shaping; scaling efforts can reveal misalignments that require returning to earlier questions about identity, target audience, or the role of AI in the business. Changes in the AI tool landscape can also reopen the imaginative field by making previously infeasible ideas suddenly realistic for a single founder.

Over time, experienced AI-augmented solopreneurs often develop a meta-capability; the skill of moving fluidly between imagining, testing, and systematizing---and of reconfiguring their human--AI setup---in response to changes in their market, technology, and personal circumstances. In the next section, we discuss the implications of this framework for empirical research and practical support interventions aimed at individuals seeking to turn their imagination into a sustainable, AI-enabled solo business.

\begin{figure*}[t]
\centering
\includegraphics[width=\linewidth]{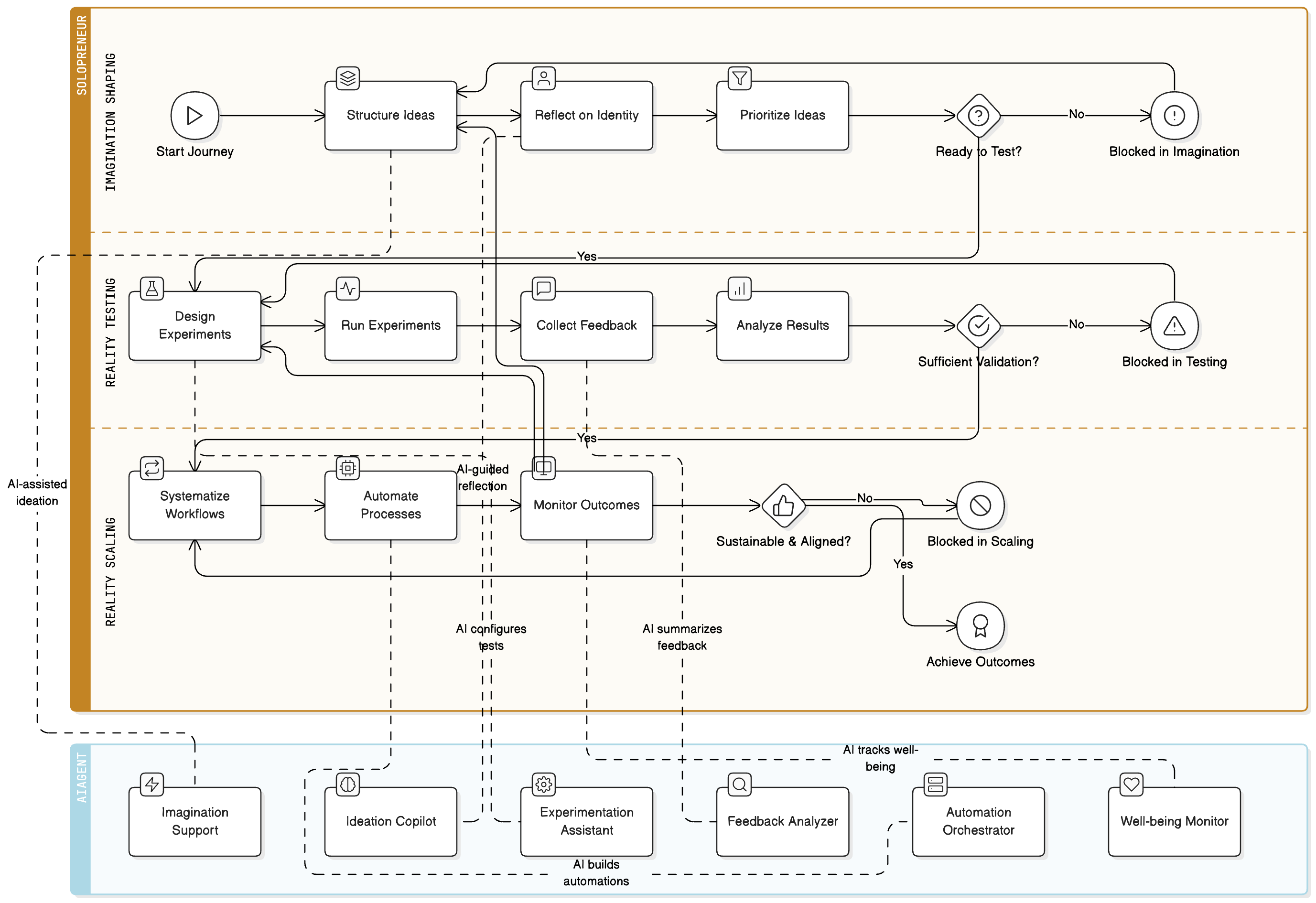}
\caption{Operational journey of an AI-augmented solopreneur across imagination shaping, reality testing, and reality scaling, with illustrative AI agent roles that act as digital co-founders at each step.}
\label{fig:journey-agents}
\end{figure*}

\section{Implications for Research and Practice}
\label{sec:implications}

Our framework has implications not only for how AI-augmented solo business is understood conceptually, but also for how it can be studied empirically and supported in practice. In this section, we first outline directions for research, with a particular emphasis on measurement of staged transformation and human--AI configurations, and then derive design principles for tools and agents aimed at solopreneurs. Figure~\ref{fig:journey-agents} translates the three-stage framework into an operational workflow, showing the concrete activities solo founders perform and the corresponding AI agents acting as “digital co-founders’’ at each step.

\subsection{Research Implications: AI-augmented Solo Business as a Process of Staged Transformation}

By framing solo business as three interrelated stages---imagination shaping, reality testing, and reality scaling---we encourage a shift from static categorizations of self-employment to process-oriented inquiry. Incorporating AI agents into this view further invites attention to how human and machine agency are distributed and reconfigured across stages. This has several implications.

\subsubsection{Longitudinal, Stage- and AI-sensitive Designs}
Existing work often treats solo workers as a single snapshot (e.g., current income, status, or segment). Our model invites longitudinal studies that track founders as they:
\begin{itemize}
    \item move from idea-heavy but action-poor phases to active experimentation, sometimes triggered by the adoption of particular AI tools;
    \item learn from early failures or partial successes, including misconfigured or over-automated AI workflows; and
    \item transition from episodic projects to routinized, sustainable solo businesses supported by semi-autonomous agents.
\end{itemize}
Such designs can reveal when and how movement between stages occurs, and which factors (skills, AI literacy, support, motivations) accelerate or inhibit these transitions.

\subsubsection{Differentiation within Solo Self-employment}
The framework provides a lens to distinguish between different profiles of solopreneurs not only by sector or necessity/opportunity status, but by process patterns and human--AI configurations, such as:
\begin{itemize}
    \item those who remain trapped in imagination shaping, including founders who use AI only for ideation but rarely test;
    \item those who experiment intensively (often with AI-boosted content and prototypes) but struggle to consolidate and scale into stable systems; and
    \item those who cyclically re-imagine and re-scale as their life context or AI tool landscape changes, reconfiguring agents and workflows along the way.
\end{itemize}
This enables richer typologies based on \emph{how} solo businesses are built and automated, not just what they are.

\subsubsection{Linking process, AI Use, and Outcomes}
The three stages can be analytically linked to economic outcomes (revenue, stability, and growth), psychological outcomes (well-being, identity coherence, and autonomy satisfaction), and relational outcomes (quality of client relationships, perceived authenticity, and network embeddedness). Researchers can examine, for example, whether robust imagination-shaping practices predict less harmful pivoting, or whether effective reality scaling with AI support correlates with long-term well-being rather than burnout or techno-overload. The framework also enables questions about when AI use is performance-enhancing versus identity-threatening or dependency-creating.

\subsection{Measurement Approaches: Operationalizing Stages and Human--AI Configurations}

Our conceptualization suggests several measurement avenues for capturing the mechanisms, resources, psychological factors, and AI-related dimensions associated with each stage.

\subsubsection{Stage Engagement and Progression}
Researchers can develop scales or structured interviews to assess:
\begin{itemize}
    \item the degree of engagement with each stage (e.g., frequency and depth of idea structuring, number and nature of experiments, and presence of routines and systems);
    \item the dominant stage at a given time (where time, energy, and attention are mostly spent);
    \item perceived blockages (e.g., ``I have many ideas but do not test them'', ``I test, but nothing becomes stable'', and ``I have a system but no time for new ideas'').
\end{itemize}
This allows empirical testing of the idea that successful AI-augmented solopreneurs move iteratively across all three stages, rather than over-investing in just one.

\subsubsection{Mechanism-specific Indicators}
Within each stage, concrete mechanisms can be operationalized, for instance:
\begin{itemize}
    \item \emph{Imagination shaping:} clarity of value proposition; perceived alignment between business idea, identity, and preferred AI role; diversity of considered options; use of reflective/structuring practices and AI-assisted ideation or market scanning.
    \item \emph{Reality testing:} number and type of experiments run; use of pre-selling, prototypes, or trial AI agents; sources and diversity of feedback; explicit decision rules (persevere/pivot/abandon); degree of AI involvement in experiment execution and analysis.
    \item \emph{Reality scaling:} presence of documented processes; degree and type of automation; complexity of the AI/tool stack; use of external tools/freelancers; predictability of revenue; boundary-setting practices.
\end{itemize}
These indicators could be captured through self-report, platform or tool usage logs (e.g., automation triggers, agent runs), or qualitative coding of founder narratives.

\subsubsection{Psychological, Relational, and AI-specific Constructs}
Our framework highlights the importance of identity, emotion, social context, and attitudes toward AI. Relevant constructs include:
\begin{itemize}
    \item identity coherence and identity work (how founders narrate who they are becoming through the business and their AI setup);
    \item learning orientation and resilience (responses to failed experiments, including AI failures or hallucinations);
    \item perceived emotional risk in visibility and experimentation, potentially shaped by AI-produced content;
    \item trust in AI, perceived authenticity of AI-assisted outputs, and clarity of human--AI role boundaries;
    \item network resourcefulness (ability to leverage communities, mentors, early clients, and even shared AI tools as co-creators).
\end{itemize}
Existing scales from entrepreneurship, career, organizational psychology, and human--AI interaction can be adapted and combined with stage-specific measures.

\subsubsection{Mixed-methods, Digital Trace Data, and Experience Sampling}
Because solo entrepreneurship is deeply intertwined with daily life and digital tools, experience sampling methods (short in-situ surveys), mixed-method designs, and analysis of digital trace data can capture how founders move between stages even within a week---e.g., ideating with an AI assistant on Monday, testing a micro-campaign on Wednesday, and systematizing a workflow on Friday. This can surface patterns of fluctuation and AI use that cross-sectional surveys miss.

\subsection{Practical Implications: Designing Support for AI-enabled Solo Founders}

Beyond research, the framework suggests concrete ways to design tools, programs, and policies that better fit the realities of AI-augmented one-person businesses. We focus here on design principles for tools and interventions, with particular attention to AI agents that align with each stage.

\subsubsection{Support for Imagination Shaping}

Tools and programs targeting this stage should help founders structure, prioritize, and emotionally ``safe-guard'' their ideas, while using AI to expand possibilities without overwhelming or displacing human judgment:
\begin{itemize}
    \item \textbf{Idea-structuring interfaces and AI co-pilots} that guide users from personal motivations and strengths to clearer problem--solution hypotheses (e.g., prompts that explicitly connect ``what matters to me'' with ``who could benefit,'' accompanied by AI-generated candidate value propositions).
    \item \textbf{Portfolio and prioritization aids} where AI helps compare multiple ideas using simple, founder-centric criteria (fit with identity, feasible next step, affordable loss, desired role of AI), while keeping the final choice transparent and firmly in human hands.
    \item \textbf{Identity-safe experimentation commitments} generated or refined by AI agents, such as templates that help founders define very small, bounded tests (e.g., one post, three conversations, and one micro-prototype) to bridge imagination and action without feeling that a whole identity or AI strategy is on the line.
\end{itemize}
Educational programs could incorporate guided reflection, narrative exercises, and AI-assisted coaching that help solopreneurs see patterns in their ideas and articulate them in market-relevant language.

\subsubsection{Support for Reality Testing}

For reality testing, effective support reduces the friction, cost, and emotional burden of running experiments, while keeping AI-mediated experimentation understandable and controllable:
\begin{itemize}
    \item \textbf{Experiment libraries and wizards} that suggest concrete, low-cost tests tailored to the type of business (e.g., coaches, designers, and SaaS builders) and to the founder’s risk tolerance, with AI agents helping to configure assets (copy, visuals, and scripts) for each test.
    \item \textbf{Feedback-collection and analysis tools} that integrate simple interview scripts, survey templates, or structured note-taking, and then use AI to summarize responses, extract themes, and highlight contradictions---without hiding raw data.
    \item \textbf{Learning dashboards and experiment trackers} where AI keeps a memory of what has been tried, visualizes evidence for or against each hypothesis, and suggests next experiments or possible pivots, reducing the sense of randomness and repetition.
\end{itemize}
Group-based programs (accelerators, masterminds, and incubators for solopreneurs) can emphasize shared debriefing of experiments, turning potentially discouraging outcomes---including AI misfires---into collective learning.

\subsubsection{Support for Reality Scaling}

At the scaling stage, the main challenge is not just ``growth'' but sustainable, personally aligned consolidation, often mediated by automation and agents:
\begin{itemize}
    \item \textbf{System-building tools} that help solopreneurs convert recurring activities into checklists, templates, and workflows, with AI agents proposing automations but explaining them in understandable steps and allowing easy rollback.
    \item \textbf{Orchestrated automation and agent layers} that connect scheduling, payments, communication, and content tools into a coherent ``solo operating system.'' Here, multi-agent setups can be designed so that one ``orchestrator'' agent coordinates specialized agents (e.g., marketing, operations, and analytics) under human oversight.
    \item \textbf{Boundary-setting and capacity-management features}, such as capacity indicators (e.g., suggested maximum number of clients or projects given current systems), working-hour guards, and AI prompts that help founders decline misaligned opportunities or resist always-on work norms that automation can exacerbate.
\end{itemize}
Training and coaching can focus on designing a solo business that fits a desired life architecture and a consciously chosen AI role, rather than defaulting to growth norms borrowed from team-based startups or to maximal automation.

\subsection{Integrating Support Across Stages}

Finally, our framework suggests value in integrated support ecosystems that recognize movement across stages rather than treating ideation, validation, and growth as separate worlds. Tools might provide \emph{stage-awareness} (e.g., quick self-assessments) and adapt their interfaces and AI guidance accordingly; offering more reflective, exploratory prompts during imagination shaping; experiment templates during reality testing; and workflow/automation suggestions during reality scaling. Agents themselves can be designed with ``mode switches'' that reflect these stages. Programs could explicitly cycle participants through imagination shaping, reality testing, and reality scaling modules, normalizing the idea that returning to earlier stages is a sign of learning, not failure. AI tools embedded in these programs can give continuity across stages by keeping a structured history of ideas, experiments, and systems. Policy initiatives and institutional support (e.g., university incubators and public entrepreneurship programs) can design solopreneur-specific tracks that respect the identity, resource, and lifestyle realities of one-person businesses, while also addressing issues such as AI literacy, responsible AI use, and access to safe experimentation environments.

Taken together, these implications highlight that supporting solo business is not just a matter of generic ``entrepreneurship training'' or generic AI adoption. It requires attention to how individual founders structure their imagination, conduct low-cost experiments, and build minimal yet robust human--AI systems that allow a one-person venture to be both economically viable and psychologically sustainable.

\section{Conclusion}
\label{sec:conclusion}

This paper has conceptualized AI-augmented solo business as a distinct imagination-to-reality context, rather than a scaled-down version of conventional startups or a residual category in the labour market. Building on four streams of literature --- solo self-employment and one-person enterprises, entrepreneurial imagination and creativity, process-based opportunity development, and AI in entrepreneurship --- we have argued that the rise of agentic AI systems fundamentally reshapes what one person can imagine, test, and sustain as an economically viable venture. We proposed a three-stage framework that articulates how AI-augmented solopreneurs move from inner vision to sustainable business reality: \emph{imagination shaping}, \emph{reality testing}, and \emph{reality scaling}. Each stage is characterized by specific inputs, mechanisms, resources (including AI agents and digital infrastructures), and psychological factors (including identity, emotional safety, and resilience). Imagination shaping highlights how personally meaningful, AI-expanded possibility spaces are translated into explicit, testable value hypotheses under solo-specific constraints. Reality testing reframes classic entrepreneurial experimentation in terms of micro- and nano-experiments that fit one person’s time, emotional bandwidth, and resource limits, increasingly supported by AI for design, execution, and analysis. Reality scaling shows how stable solo businesses emerge not from organizational layering, but from simple systems, routines, and orchestrated human--AI workflows that extend capacity without requiring employees. Across these stages, we articulated four interrelated contributions. First, we positioned AI-augmented solo business as a theoretically distinct context that complements structural and policy-oriented analyses with micro-process views (addressing Gap~1). Second, we embedded entrepreneurial imagination and creativity in the lived realities of solopreneurs, emphasizing identity, emotional regulation, and human--AI collaboration (addressing Gap~2). Third, we adapted effectuation, lean startup, and design thinking to one-person ventures operating with agentic AI, reorganizing them into the three-stage framework (addressing Gap~3). Fourth, we bridged macro descriptions of the solo economy with micro mechanisms and AI design implications, providing a basis for future empirical work and for the design of tools and ecosystems that better support one-person businesses (addressing Gap~4).

Our framework opens several avenues for empirical research and theory development on AI-enabled solopreneurship and digital co-founders:

\subsubsection{Empirical Validation of the Three-stage Model}
A first line of inquiry is to operationalize and empirically test the imagination shaping, reality testing, and reality scaling stages. Longitudinal and mixed-method studies could examine whether and how solopreneurs move across these stages over time, how often they loop back, and which trajectories are associated with different outcomes (e.g., income stability, growth, well-being, and identity coherence). Such work can also investigate whether distinct process profiles exist (e.g., ``permanent imaginers'', ``chronic testers'', and ``over-scalers'') and how AI adoption influences these trajectories.

\subsubsection{Measurement of Mechanisms, Psychological Factors, and AI Configurations}
A second direction concerns measurement. Future studies can develop and validate instruments that capture key mechanisms within each stage (e.g., clarity of value hypotheses, density and quality of experiments, and robustness of routines and systems), as well as psychological variables (e.g., fear of visibility, resilience, learning orientation, trust in AI, and perceived authenticity of AI-assisted work). Combining self-report, qualitative data, and digital trace data from AI tools and platforms could yield richer models of human--AI configurations and their impact on both performance and well-being.

\subsubsection{Comparative Studies Across Solo Segments and Contexts}
Third, the framework can be used to compare different segments of solo workers and solopreneurs; necessity- versus opportunity-driven, creative versus technical sectors, early-career versus late-career, or low- versus high-AI-adoption profiles. Cross-country or cross-institutional comparisons could illuminate how regulatory regimes, social protection systems, and ecosystem support (incubators, platforms, and communities) shape access to AI tools, experimentation spaces, and sustainable scaling paths for one-person ventures.

\subsubsection{Design and Evaluation of AI Agents and Tool Ecosystems}
Fourth, our stage-based perspective suggests design requirements for AI agents and tool ecosystems that support solopreneurs as \emph{digital co-founders}. Future research can explore how different agent roles (e.g., ideation co-pilot, experiment architect, and systems orchestrator) affect founders’ decision-making, learning, and sense of agency. Experimental and field studies can evaluate the impact of stage-aware tools --- systems that adapt their guidance and automation to whether the founder is primarily imagining, testing, or scaling --- on business outcomes and psychological sustainability.

\subsubsection{Ethical, Identity, and Policy Implications}
Finally, AI-augmented solo business raises ethical, identity, and policy questions that merit dedicated study. At the micro level, research is needed on how reliance on AI agents affects founders’ sense of authenticity, ownership, and responsibility, and how they communicate AI involvement to clients. At the macro level, policymakers and institutions must grapple with how to support a growing population of AI-intensive solopreneurs in terms of social protection, lifelong learning, and responsible AI use. Our framework offers a vocabulary for analyzing how interventions at the ecosystem level (e.g., access to tools, experimentation infrastructure, or stage-specific training) map onto the micro-processes by which individuals turn imagination into economic reality.

\end{document}